\documentclass[12pt, prd]{revtex4}
\usepackage{graphicx,epsfig}
\input{epsf.tex}
\usepackage{amsmath}
\usepackage{amssymb}

\usepackage[utf8]{inputenc}

\begin{document}

\title{The Nobel prizes in physics for astrophysics and gravitation and
the Nobel prize for black holes: Past, present, and future}
\author{Jos\'{e} P. S. Lemos}
\affiliation{Centro de Astrof\'isica e Gravita\c{c}\~ao - CENTRA,
Departamento de F\'isica, Instituto Superior T\'{e}cnico - IST,
Universidade de Lisboa - UL, Avenida Rovisco Pais 1, 1049-001, Lisboa,
Portugal, Electronic address: joselemos@tecnico.ulisboa.pt}



\begin{abstract}
We analyze the Nobel prizes in physics for astrophysics and gravitation
since the establishment of the prize and highlight the 2020 Nobel prize
for black holes. In addition, we comment on the names that could have
received the prize in astrophysics and gravitation, and draw attention
to the individuals who made outstanding contributions to black hole
physics and astrophysics and should be mentioned as possible and
deserved recipients of the prize.  We speculate about the branches of
research in astrophysics and gravitation, with an emphasis on the
latter, that can be contemplated in the future with a Nobel prize.

\end{abstract}
\maketitle

\section{Introduction}

The Nobel prize in physics, awarded by the Royal Swedish Academy of
Sciences to exceptional scientists and works, was inaugurated in 1901,
having in that year been attributed to Röngten, a German physicist,
for the ``the discovery of the remarkable rays subsequently named
after him''.  
by the Nobel 
electromagnetic radiation of high energy. The prize is conceded
annually, and up to now several areas of physics had the privilege of
being graced with this magnificent distinction.

The main criterion for the Nobel prizes in physics is that the
scientific works and achievements to be considered for election have
been time tested. In practice, this means that only established
experimental works that led to the progress of physics and pioneering
theoretical works that have been experimentally verified can aspire to
the prize.  Pure speculation by itself, for all the interest that it
can have, does not adhere to this criterion and is simply not
considered.

The process of selecting the candidates is judicious. It involves the
Royal Swedish Academy of Sciences that elects the Nobel committee for
physics, composed by five people, that in turn receives nomination
proposals from previous Nobel prize in physics receivers and from
other scientists. After selecting ten nomination proposals, the
committee submits those to the Academy that, after the necessary
meetings, votes in the prize winners of that year. At most there are
three winners per year, a strict rule that the committee and the
Academy adhere to.

This combination of confirmation of the work along a sufficient long
time, of the hardworking and thorough laboring of the committee and
the Academy, and of admitting at most three recipients, is very
powerful and make the Nobel prize in physics, by farm the most
important prize in physics.  The Nobel prize in physics, beyond being
a prize for exceptional physicists, is a prize for physics as a
science, and the individual that receives it has that blessed
felicity.

Within several areas of physics, we are interested in the areas of
astrophysics and gravitation, which are two areas closely interlinked.
An instance of this link is cosmology, the science of the Universe,
that uses astrophysics to explain the physics of the relevant
phenomena and works with gravitation to elucidate the global dynamics
of the spacetime. Another instance of this interlink between the two
areas is the black hole, an object with its existence established
through astrophysical observations and also a product of pure
speculation within gravitation, in particular within the general
theory of relativity.

We will describe and analyze to whom and for which works the Nobel
prize in astrophysics and gravitation was ascribed, including and
highlighting the 2020 distinction to black holes. We will comment also
on possible recipients, that for one reason or another were not
honored with that consecrated prize of the Swedish Academy. In
addition, we will speculate on possible branches of astrophysics and
gravitation that can obtain in the future a Nobel prize.

\section{The Nobel prize in physics for astrophysics and
gravitation from its creation to 2019}\label{ag}
\subsection{The recipients}

Astrophysics and gravitation are two vast branches of physics
intimately connected since Newton discovered that gravitation is
universal, linking conclusively the heavens and the Earth, and
revealing a unified Universe with everything obeying the same laws of
physics. The inaugural Nobel Prize attributed to Röngten for the
discovery of X rays is certainly a prize neither to astrophysics nor
to gravitation, although X rays have turned out to be an extremely
important observational window in astrophysics.  The Nobel prizes
attributed directly to astrophysics and gravitation are several. We
review them here, providing the date of attribution of the prize, the
name of the recipient, and a summary of the text provided by the Nobel
committee of the Royal Swedish Academy of Sciences, while weaving in
some commentary on the corresponding and related accomplished works.

1921 Einstein - ``For his services to theoretical physics, and
especially for his discovery of the law of the photoelectric effect''.
It is common practice to state that Einstein got the prize for the
theoretical explanation of the photoelectric effect, but the
announcement of the Nobel committee states much more than that, it
explicitly refers to the services rendered to theoretical physics.
Einstein's services to theoretical physics are very many, and clearly
we can include in those services the creation of the special and
general relativity.  General relativity is the theory of gravitation
that succeeded Newton's theory of gravitation and thus this prize is
also a prize for gravitation.

1967 Bethe - ``For his contributions to the theory of nuclear
reactions, especially his discoveries concerning the energy production
in stars''.  Bethe was a German physicist that settled in the United
States. A question raised by Eddington, a renowned English
astrophysicist, was the need to solve the problem of energy generation
in the center of a star, the Sun being an example. It was admitted
that the energy should come from nuclear reactions, possibly from the
transformation of four protons into a helium nucleus. The complicated
details of those reactions that occur at the center of the stars were
devised in 1939 by Bethe, he was by then in Cornell. This is a Nobel
prize for both astrophysics and nuclear physics.

1974 Ryle; Hewish - ``For their pioneering research in radio
astrophysics. To Ryle for his observations and inventions, in
particular of the aperture synthesis technique, and to Hewish for his
decisive role in the discovery of pulsars''. Ryle was an English
astrophysicist based in Cambridge who developed a system of antennas
to perform observations in radio waves. He showed
around 1960 that distant
galaxies, and thus younger galaxies, were different from nearby
galaxies, and thus older galaxies, having this fact on one hand helped
to corroborate the veracity of a universe in expansion and so in
transformation, on the other hand, having been one of the first
arguments of weight to demonstrate that the cosmological theory of a
steady state universe, proposed by Bondi, Gold, and Hoyle, theoretical
astrophysicists also based in Cambridge, was wrong, or putting it
better, was a theory not compatible with our observed Universe.
Hewish, an English astrophysicist also based in Cambridge, working
also in radioastronomy, discovered in 1967 the first pulsar through
the detection of fast wave pulses in radiofrequency coming from the
sky. His PhD student Jocelyn Bell was in command of the instruments
when the signal arrived, and reckoning, upon analyzing the registers,
that was a new kind of signal and possibly an important one,
immediately showed it to Hewish and the group. Soon after, it was
shown by Gold, one of the Cambridge physicist of the steady state
universe that was at the time in Cornell, and by other physicists,
that pulsars were neutron stars in rapid rotation, neutron stars being
stars of about one solar mass within a radius of 10 Km constituted
essentially from neutrons. Neutron stars had been speculated to exist
by Zwicky, a Swiss astrophysicist working at Caltech, in 1934, as
stars that would remain in the center of a supernova explosion. This
is a Nobel for astrophysics.

1978 Penzias and Wilson - ``For their discovery of cosmic microwave
background radiation''. Penzias, born in Germany and based in the USA,
and Wilson, a Caltech-trained American physicist, had done their PhDs
in microwave physics and radioastronomy and worked in the radio
antennas of Bell Laboratories, in New Jersey. In 1965 they detected
the cosmic microwave background radiation, which has a temperature of
2.7 K. This cosmic radiation had been predicted by Alpher and Herman,
American physicists, disciples of Gamow, a Russian physicist who had
settled in the United States of America. The detection by the
fortunate Penzias and Wilson was by chance, they did not know the
origin of the detected radiation. By coincidence, Dicke, an American
physicist competent both in theoretical and experimental physics, and
his group that included Peebles in Princeton, neighbors of the Bell
Labs, were looking for that detection and interpreted immediately the
detected radiation as being the cosmic microwave background
radiation. Thus, this detection, that could have been lost without any
explanation, was clarified immediately and with that, the big bang as
the standard model of the Universe was confirmed convincingly. One can
argue that it is strange that Penzias and Wilson have received the
prize for the discovery of a phenomenon that they did not yet
understand, but the fact is that it is a discovery of first magnitude
done by the two and it is within the criteria of the Nobel prize. The
situation of the inaugural prize is identical. Röngten had no idea
what rays he had discovered, only later it was inferred that it was
high energy electromagnetic radiation, and nobody, or almost nobody,
knows who has made that inference. In any case, beginning in 2023, the
1978 Nobel prize archives will in principle be available to be
consulted and we can then understand the arguments advanced for this
recommendation. This is a Nobel prize for astrophysics and
gravitation, more precisely to cosmology which is an instance of link
between these two scientific areas. This discovery of Penzias an
Wilson constituted half of the 1978 prize; the other half went to
Kapitsa, an influential soviet physicist working in low temperature
phenomena.

1983 Chandrasekhar; Fowler - To Chandrasekhar ``for his theoretical
studies of the physical processes of importance to the structure and
evolution of the stars'' and to Fowler ''for his theoretical and
experimental studies of the nuclear reactions of importance in the
formation of the chemical elements in the universe''. Chandrasekhar,
was born in India, and went to Cambridge in his youth, where in the
1930s developed a remarkable work on the structure of white dwarfs,
having discovered with precision what is the maximum mass of those
stars, specifically, the mass above which gravitational collapse is
inevitable.  This mass is 1.4 solar masses. He had many other
important works in astrophysics and gravitation. In one of them, in
1942, in collaboration with M\'ario Shenberg, a Brazilian physicist,
when Shenberg was in Chicago where Chandrasekhar had installed
himself, it was found a maximal mass to the central part in isothermal
equilibrium of a red giant, called Schönberg-Chandrasekhar limit,
above which the very center shrinks under its own weight, increasing
its own temperature and beginning the burning of the existing
helium. Fowler, an American physicist of Caltech, was a specialist in
theoretical and experimental nuclear physics. In a paper in 1957 by
the English couple Burbidge and Burbidge, Fowler, and Hoyle, known as
the B$^2$FH, it was shown that all the chemical elements heavier than
helium are generated in the center of the stars. This is a Nobel prize
for astrophysics and gravitation from Chandrasekhar's side and it is a
Nobel prize for astrophysics and nuclear physics from Fowler's side.

1993 Hulse and Taylor - ``For the discovery of a new type of pulsar, a
discovery that has opened up new possibilities for the study of
gravitation''. Taylor, an American astrophysicist from Princeton, and
Hulse, American and PhD student at the time, discovered in 1974 a
binary pulsar, constituted by a pulsar and a neutron star, in close
proximity and in a conjoint orbit. This binary system of compact stars
enabled the deduction of its own physical proprieties. In particular,
the decrease of the orbital radius with time of the binary system is
in accord with the general relativistic prediction taking into account
the energy loss by gravitational radiation given by the theory. It is
the first proof of the existence of gravitational waves, albeit
indirect. This is a Nobel prize for gravitation through astrophysics.

2002 Davis, Koshiba; Giacconi – To Davis and Koshiba ``for pioneering
contributions to astrophysics, in particular for the detection of
cosmic neutrinos'' and to Giacconi ``for pioneering contributions to
astrophysics, which have led to the discovery of cosmic X-ray
sources''.  Davis, an American physicist, and Koshiba, a Japanese
physicist, prepared sophisticated experiments in the 1960s in which
neutrinos from the center of the Sun were detected for the first time,
giving a crucial direct proof of the existence of nuclear reactions
there. However, the flux of neutrinos received was a third of the
calculated flux from the theory of stellar structure and the theory of
elementary particles.  This discrepancy opened a new line of research
in neutrino physics and it gave rise to the experimental discovery
that neutrinos have mass, although extremely small, and in flight they
oscillate changing the identities in-between them, the original
detectors being only apt to capture only one of the three neutrino
identities. For the solution of this problem, Kajita, a Japanese
physicist, and MacDonald, an American physicist, received the physics
Nobel prize in 2015.  Giacconi, an Italian physicist, that worked in
several projects in Europe and in the United States of America, was
one of the precursors in the construction of X ray detectors, or
telescopes, the same X rays that had been discovered by Röngten. These
detectors have to be in orbit to avoid the Earth's atmosphere which is
opaque to X rays. An explorer satellite with an X ray detector on
board, the Uhuru, was launched to space in 1970.  One of the X ray
sources, called Cygnus X1, that had been identified in the
constellation Cygnus in 1964 by a predecessor satellite, was the
target of special observations by the Uhuru satellite.  The X ray
intensity coming from the source suffered quick variations which, due
to the finiteness of the speed of the light, indicated a very compact
object.  Additional studies showed that the source was composed of a
binary system, with one of the stars being a blue giant of about 30
solar masses, and the other being a highly compact object of about 20
solar masses, which could only be a black hole, the first stellar
black hole candidate to be identified. The observed X rays are
generated by the collision of the matter incoming from the blue giant
agianst the matter of a disk revolving around the black hole. This is
a Nobel prize for astrophysics and elementary particle physics from
the side of Davis and Koshiba, and it is a Nobel for astrophysics,
gravitation and other areas of physics from the side of Giacconi.

2006 Mather and Smoot – ``For their discovery of the blackbody form
and anisotropy of the cosmic microwave background radiation''. Mather,
an American physicist, specialist in instrumentation and cosmic
microwave background radiation, and Smoot, an American physicist,
specialist in instrumentation, particle physics and astrophysics, were
in command of the science of the COBE satellite, acronym for Cosmic
Background Explorer.  After the spectacular detection in unexpected
circumstances of the cosmic microwave background radiation in 1965,
the next task was to understand the detailed physics of that
radiation. In the primordial universe, radiation and matter were in
thermal equilibrium, this radiation having a characteristic black body
spectrum, until, due to the expansion of the Universe, the radiation
decoupled from the matter and the corresponding spectrum having been
congealed at this very time. For this reason, the detected radiation
should have a blackbody spectrum, the signature of the initial thermal
equilibrium. In addition, in the primordial universe there were matter
density fluctuations and, due to the coupling between matter and
radiation at that time, this implied that there were fluctuations in
the temperature of the radiation. These fluctuations supposedly
originated the galaxies. At the moment of decoupling, those
fluctuations were imprinted in the temperature of the radiation, the
temperature of a point in the sky should have a different temperature
from a neighbor point in the sky.  Mather, in charge of the form of
the spectrum of the radiation obtained a perfect blackbody form for
it, and Smoot, in charge for the measurements of the temperature
fluctuations, obtained fluctuations of one part in $10^5$, which was
the expected value, and that many others before him tried to find but
did not manage. This is a Nobel prize for astrophysics and
gravitation, more precisely, to cosmology which is an instance of a
link between these two scientific areas.

2011 Perlmutter, Schmidt and Riess – ``For the discovery of the
accelerating expansion of the Universe through observations of distant
supernovae''. Perlmutter is an American physicist based in Berkeley,
Schmidt is an American physicist based in Australia, and Riess is an
American physicist based in the John Hopkins University.  Since the
discovery by Hubble in 1929 of the linear relation of galaxy redshifts
as a function of their distance that one was after to know whether
this linearity would change with the distance into a nonlinear
relation, and if it did, to which side of the line the law would
change. If it changed to one side it would mean that the Universe was
decelerating, and if it changed to the other side it would mean the
Universe was accelerating.  In the years 1990s, Perlmutter understood
that the supernovas type IA, stars that explode just after they attain
the Chandrasekhar limit by extracting matter from the companion star,
if well used, would serve as standard candles, and because of that
they would be excellent gauges for the far away distances since their
brightness is extremely high.  In 1988, after a great effort to obtain
telescope time in all possibles places of the planet, Perlmutter
announced that the expansion of the Universe is accelerated.  Schmidt
and Riess, in a similar and parallel project, discovered concomitantly
this accelerated expansion of the Universe. A physical explanation of
this accelerated expansion requires a cosmological constant, or some
other form of dark energy possibly associated to the vacuum energy, a
subject that has stimulated an intense investigation in the areas of
gravitation, quantum field theory, and elementary particle theory.
This latter has been of enormous importance in cosmology, in the
explanation of the physical phenomena of the primordial universe,
after it was understood in 1973, from the works of the American
physicists Politzer, Gross, and Wilczek, contemplated with the Nobel
prize of 2004, that the strong interaction is asymptotically free, and
so the three fundamental interactions, namely, the electromagnetic,
the weak, and the strong interactions, can be united in a grand
unified theory at a high energy scale, a little smaller than the
Planck energy scale, the larger elementary energy scale possible.
This is a Nobel prize for astrophysics and gravitation, more
precisely, to cosmology which is an instance of a link between these
two scientific areas.

2017 Weiss, Barish, and Thorne - ``For decisive contributions to the
LIGO detector and the observation of gravitational waves''. Weiss,
born in Germany and settled in the United States of America,
specifically at MIT, Barish, American particle physicist familiar with
large science collaborations, and Thorne, an American physicist from
Caltech and Wheeler's student in Princeton, headed the LIGO
project. LIGO is an acronym for Laser Interferometer
Gravitational-Wave Observatory.  Weiss in the 1960s realized that a
laser interferometer would be the perfect apparatus to detect
gravitational waves and constructed small prototypes in such a way to
refine its functioning. Thorne, specialist in black holes and
gravitational waves, developed the theory of gravitational waves in a
systematic way and in a conjoint effort with Weiss created LIGO. At
the end of the 1990s, Barish became the new LIGO director, supervising
the construction of the two interferometers, one in the state of
Washington, the other in the state of Louisiana, separated by 3
thousand Kms. On September 14, 2015, the two laser interferometers
vibrated in the same manner with an interval of approximately one
hundredth of a second as a consequence of the passage of a
gravitational wave through the Earth.  The data gathered by the
apparatuses showed that the wave had been generated by a collision, at
cosmological distances, between two black holes of about 30 solar
masses each. It was the first time humankind detected gravitational
waves. Nowadays, one does astronomy and astrophysics with
gravitational waves. This is a Nobel prize for gravitation.

2019 Peebles; Mayor and Queloz - ``For contributions to our
understanding of the evolution of the universe and Earth's place in
the cosmos with half of the prize going to James Peebles for
theoretical discoveries in physical cosmology and the other half going
to Michel Mayor and Didier Queloz for the discovery of an exoplanet
orbiting a solar-type star''.  Peebles, born in Canada and settled in
Princeton, belonged to Dicke's cosmology original group, being one of
the authors of the 1965 paper that accompanies the paper by Penzias
and Wilson.  Peebles developed physical cosmology in a rigorous
way. Among the several important contributions, he showed the physics
that one could extract from the cosmic background radiation and the
importance of the dark matter in galaxy formation.  Mayor and Queloz,
two Swiss astrophysicists based in Geneva, discovered for the first
time a planet outside the solar system in a star of the Sun's
type. Nowadays about five thousand exoplanets are known and there are
about the same amount of candidates waiting for validation. This is a
Nobel for astrophysics and gravitation, more precisely to cosmology
that is an example of the interlink between these two scientific areas
from Peebles side and is a Nobel for astrophysics from Mayor and
Queloz side.

\subsection{The ones who could have won the prize}
\label{osque}

It is of interest to indicate and comment on those physicists that could
have been honored with the Nobel prize in physics in astrophysics and
gravitation, and for some reason the Nobel committee and the Royal
Swedish Academy of Sciences did not place them in the maximum priority
list.

Eddington, the great English astrophysicist from Cambridge, was the
first to understand around 1920 the interior of stars through a simple
model that permits the calculation of the temperature and the density in
the center of a star.  He also recognized that the energy generated in
the interior of a star had to be subatomic, i.e., it had to be nuclear
energy, and this was confirmed when Bethe found the nuclear reactions at
the center of star that did indeed generate the necessary energy, having
received the Nobel prize for that work. Eddington gave many other
contributions, namely, to galactic physics and  galactic astronomy, to
general relativity, notably to its confirmation through the test of
light deviation in the eclipse of 1919 when he went to Principe Island,
to cosmology, to fundamental physics, and to the general understanding
of science. In the first two decades of the 20th century, works in
astrophysics were hardly, or even never, considered for the Nobel prize,
and for that reason one understands why Edddington never received it.
It would have been a Nobel prize for astrophysics.

Hubble, an American astronomer, was head of the largest telescope in
the world, located in California, the Mount Wilson telescope of 2.5
meters.  Through precise observations he was able to produce a
relation between the redshift and the distance of galaxies, and it
allowed him to announce in 1929 the law that the velocity of a galaxy
is proportional to its distance, known as the Hubble law. It was an
extraordinary feat in astronomy. Note, nevertheless, that Hubble
always stated that the relation did not mean for sure that the
Universe was expanding.  In this connection, it must be remembered
that Lema\^itre, a Belgium physicist, disciple of Eddington,
discovered between 1925 and 1927, with the observational data so far
published and in possession of a solution of general relativity that
gave an expansion universe that he himself had found, although
Friedmann, a Russian physicist from St. Petersburg, had found the same
solution three years before in 1922, the law that furnishes the
proportional relation between the galaxy velocity and its distance,
and for this reason is often called Lemaître-Hubble law.  In addition,
Lemaître was the first to propose a big bang scenario consistent with
the Universe in expansion.  It is hard to fathom why Hubble did not
receive the physics Nobel prize. The discovery of the linear relation
between the redshift and the distance of the galaxies, and the
consequent inference that that means a Universe in expansion, is one
of the greatest scientific discoveries of all times, done with a
thorough, precise, and extremely difficult observational work.
Perhaps, the Swedish Academy had doubts whether an observational work
was eligible for a Nobel prize in physics, an error that was not
repeated as the Nobel prizes for pulsars,
 cosmic microwave background radiation,
binary pulsars,
 solar
neutrinos, 
X-ray observations, nature of the cosmic microwave background radiation,
and acceleration of the Universe, testify.  By inherence, Lema\^itre
could have received the Nobel prize with Hubble, although his work was
more speculative and of difficult confirmation at the time. It would
have been a Nobel prize for cosmology.

Gamow, born in Crimea in imperial Russia, obtained his PhD in St.
Petersburg during the Soviet era, and later installed himself in the
United States of America, first in Washington DC, and afterwards in
Colorado. Gamow made decisive contributions to physics, astrophysics,
and cosmology.  In 1928, he discovered the tunneling effect in the deacy
of atomic nuclei, a generic effect and of enormous importance in quantum
phenomena. From 1942 onwards he developed a big bang model in cosmology,
in which the expanding Universe, in its first three minutes manufactured
all the known elements starting with the hydrogen, from deuterium, to
helium and to the transuranics.  It was proved later that the light
chemical elements, as deuterium, helium, and lithium, were manufactured
in the primordial universe, whereas all other heavier elements were and
are manufactured in the center of stars, according to the analysis of
Hoyle and collaborators. Alpher and Herman, disciples of Gamow, proposed
in 1948 the existence of a cosmic background radiation, a relic of that
Gamow's primordial universe. That radiation was detected 17 years later
by Penzias and Wilson, and interpreted as being of cosmic origin by
Dicke and his group.  It is a fact that Dicke and his group in the
respective inaugural papers on this subject, did not refer to the
pioneering works of Gamow, Alpher, and Herman. It is also a fact that
the radiation found was immediately interpreted correctly by Dicke.  The
Nobel committee came across many names in the theoretical side and since
the Nobel is conceded only to three people at most, those who have
detected that radiation, namely Penzias and Wilson, were preferred for
the award, leaving out Gamow, Alpher, Herman, and Dicke.  The Nobel
prize to Peebles in 2019 has also a posthumous character of a prize to
Dicke that was his supervisor and mentor at the time the cosmic
background radiation was discovered. It would have been a Nobel prize
for the theoretical side of cosmology.

Hoyle, an English astrophysicist at Cambridge, had a number of important
and decisive ideas on how chemical elements in stellar interiors are
generated.  When Hoyle was in Caltech as a visitor in 1953, he
understood that a resonance in the carbon nucleus had to occur,
otherwise carbon would not be manufactured in the stars and we would not
be here. Fowler and his group at Caltech showed experimentally the
existence of that resonance.  Hoyle was also an important author of the
1957 B$^2$FH paper in which the synthesis of all the elements in the
stars is seminally proposed; additionally, he was the author of other
papers with several collaborators in the 1960s, in which one concludes
that the majority of the existing helium would have been primordial, and
not manufactured in the stars.  Together with Cambridge physicists Bondi
and Gold, Hoyle proposed the imaginative steady state cosmology that did
not flourish because of incompatibility with observations, despite Hoyle
insisting on it to be correct up to the end. Hoyle also thought that the
DNA molecule had originated outside the Earth, perhaps in some place
within the solar system or even in some other location. He was harsh in
dealing with adversaries and for that he gained many opponents. Even
knowing the highly heterodox side of Hoyle within science, it is a
mystery why the Royal Swedish Academy of Sciences decided to give the
Nobel prize to Fowler without concomitantly awarding it to Hoyle.
Neither Fowler understood nor was Hoyle happy. This would not have been
a new Nobel prize to astrophysics, it would be within the 1983 Nobel
prize.

Shapiro, an American physicist, is emeritus professor in Harvard,
having also been a professor at MIT.  In 1964, he discovered and
proposed a fourth test of general relativity. General relativity had
been proven correct in three tests, namely Mercury's perihelion
precession, the light deflection in the gravitational field of the
Sun, and the redshift of light when climbing a gravitational
field. These tests were proposed and verified around the 1920s and
afterwards.  It was in a totally unexpected manner that the fourth
test appeared. This fourth test is the test of the delay in the radar
echo from a planet, called Shapiro delay.  When Venus is aligned with
the Sun and the Earth and is situated in the opposite side on the Sun,
the signal passes through the gravitational field of the Sun, and due
to the curvature of space and spacetime the signal suffers a delay
relatively to the propagation in a flat space, as it is the case in
Newtonian gravitation.  The test was performed and the result verified
by Shapiro himself and collaborators in 1968. It is a brilliant and
wonderful idea that confirmed once more the correctness of general
relativity, during a time in which the theory had only passed 
three earlier tests, denominated classical tests.  In addition, the
Shapiro delay has proven useful in other astrophysical systems.  One
can understand that the Academy has not awarded a Nobel prize here;
perhaps there are a number of ideas and tests of this level in other
areas of physics, with the corresponding candidates expecting to
receive the Nobel prize. It would have been a Nobel prize to
gravitation.

Jocelyn Bell, a British citizen born in Northern Ireland, is presently
rector of the University of Dundee. In 1967, she was a PhD student in
Cambridge with Hewish as supervisor. Bell was in charge of overseeing
the data that arrived at the radiotelescope and one day in 1967 noted
that there was a signal with a regular pulse with a period of about one
second. Several identical regular pulse signals appeared
soon afterward in
other parts of the sky. These sources were denominated pulsars. Given
the importance of the discovery, a paper accounting for the observation
of the first pulsar was immediately published. This 1967 article which
reports the discovery is signed by five authors, Hewish first, Bell
second. Hewish was responsible for the development of the radiotelescope
and was the head of the science to be done from the telescope. For the
discovery of the first pulsar Hewish received the Nobel prize. Bell
should also have received it conjointly, after all it was she that
discovered the first pulsar, in analyzing the registers she identified
without hesitation a nonusual radio source. The committee possibly
considered that PhD students would not have status to receive the Nobel
prize. It did not commit the same mistake again, attributing the prize
to Hulse, Taylor's student, in 1983. This would not have been a new
Nobel prize to astrophysics, it would have been part of the 1974 prize.

\section{The Nobel prize in physics 2020
for black holes}
\label{bns}

\subsection{The recipients}\label{aq}

Given that the Nobel prize for the detection of gravitational waves
was conceded in 2017 and these waves appeared from the collision of
two black holes one could dream that the attribution of a Nobel prize
for black holes would occur in the near future, and in such a case,
one could speculate on the possible winners of the award.  But dream
and speculation is one thing, and reality is another. It was with
great surprise that, on October 6, 2020, the announcement that the
Nobel prize in physics was awarded for black holes was received. The
statement of the committee for the Nobel prize in physics 2020 reads:
``The Royal Swedish Academy of Sciences has decided to award the Nobel
Prize in Physics 2020 to Roger Penrose, of University of Oxford,
United Kingdom, to Reinhard Genzel, of Max Planck Institute for
Extraterrestrial Physics, Garching, Germany and University of
California at Berkeley, USA, and to Andrea Ghez, University of
California at Los Angeles, USA.  One half goes to Roger Penrose for
the discovery that black hole formation is a robust prediction of the
general theory of relativity, and the other half goes jointly to
Reinhard Genzel and Andrea Ghez for the discovery of a supermassive
compact object at the centre of our galaxy".

Who are these Nobel prize award-winners?     

Penrose was born in 1931 in Colchester, England, obtained his PhD in
Mathematics in Cambridge and is presently the Emeritus Rouse Ball
Professor in Oxford University. He developed new mathematical
techniques in topology and differential geometry and applied them to
the study of the geometric properties of spacetimes in general and
black holes in particular. 
\begin{figure*}[h]
\centering
\includegraphics[scale=0.65]{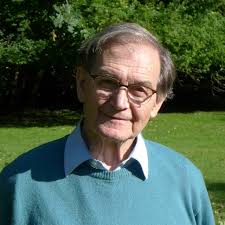}
\caption{Penrose.}
\label{rp}
\end{figure*}
He was the first to understand the essence
of a black hole.  He created in 1965 the concept of a trapped surface,
a surface from which outgoing light rays converge, and that arises
when the gravitational field interior to that surface is very
intense. With that idea he managed to show that a singularity inside
such a surface forms, presumably a singularity where the spacetime
itself ends and the laws of physics as we know them disappear, putting
an endpoint on the discussion of the inevitability of whether a black
hole and a singularity forms in gravitational collapse. For many, his
1965 paper is considered the most important contribution to general
relativity since the original formulation of the theory in 1915. But
there are more celebrated contributions. Kerr, a New Zealander
physicist working in Austin, Texas, in 1963, found an exact solution
in general relativity that corresponded to a black hole in rotation,
generalizing in an admirable fashion the solution found in 1916 by
Schwarzschild, a German physicist, to a static spacetime. Penrose
discovered in 1969 that it is possible to extract energy from a Kerr
black hole through the decaying, near the event
horizon, of a 
particle into two particles. One of the particles enters the black
hole and the other is emitted with more energy to infinity at the
expense of the black hole rotational energy. This work, of enormous
importance, triggered an entirely new line of research into the
physics of black holes that culminated with the discovery by Hawking
that black holes radiate by quantum processes. Penrose promoted
several other advances to physics and mathematics. For example, in
cosmology he perceived that in the big bang, in the first instants,
the entropy of the Universe was extremely low, as the gravitational
degrees of freedom, holders of the largest part of the entropy, were
practically frozen, although the matter degrees of freedom might have
been highly excited. This fact requires very precise initial
conditions that any cosmological theory must explain and to which the
inflationary theories in vogue cannot answer presently, unless some
form of anthropic principle in the context of a multiverse is
invoked. It is certainly a honor for Penrose to win the Nobel prize,
and it is also a privilege to the Swedish Academy to have Penrose amid
the award-winning names and for this occasion the decision merits many
congratulations.

Genzel was born in 1952 near Frankfurt, Germany. He obtained his PhD
in 1978
\begin{figure*}[h]
\centering
\includegraphics[scale=0.20]{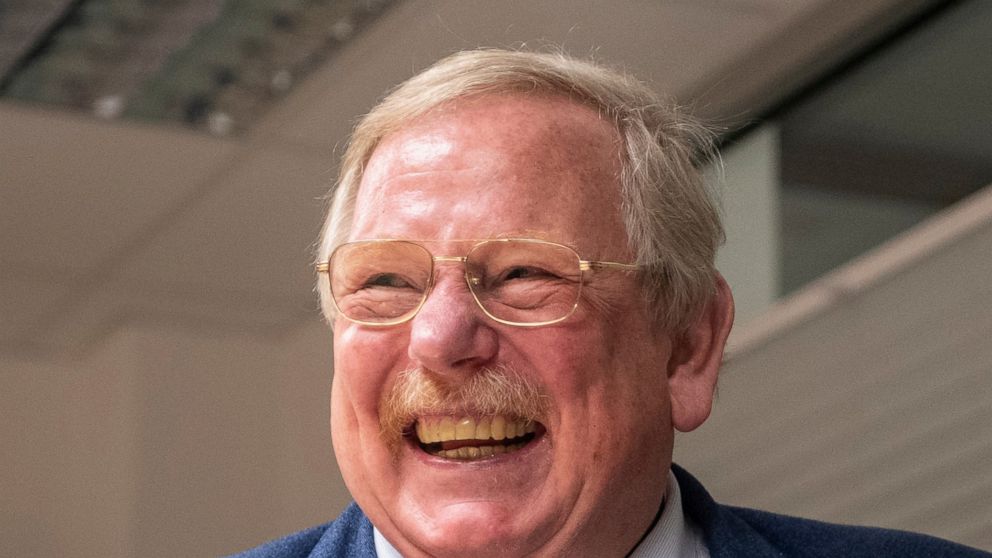}
\caption{Genzel.}
\label{rg}
\end{figure*}
from Bonn University and is now the director of the Max Planck
Institute for Extraterrestrial Physics in Garching and professor in
Berkeley. He leads the Gravity project associated to the European
Southern Observatory - ESO, that he created in the 1990s and studies
in maximum detail the stars in orbit around the very center of our
galaxy. With his group he showed that those stars have very high
velocities in their orbits, which is only compatible with the
existence of a black hole of about
4 million solar masses in the center of
our galaxy. 

Ghez was born in 1965 in New York. She obtained her PhD in 1992 from
Caltech, and
\begin{figure*}[h]
\centering
\includegraphics[scale=0.65]{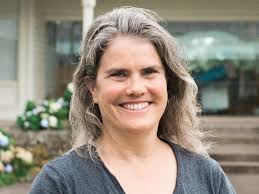}
\caption{Ghez.}
\label{agh}
\end{figure*}
is now professor in the University of California at Los Angeles. She
leads a project in the Keck telescope in Hawaii, that started at the
end of the 1990s, and that studies the kinematics of the stars in
orbit around the very center of our galaxy. Several new stars were
observed and it was also shown that these have very high velocities
only compatible with the existence of a black hole of 
about 4 mino acllion solar
masses in the center of our galaxy. Genzel's and Ghez's projects,
being totally independent, complete each other and minimize the errors
that one of the two projects may present.

\subsection{The ones who could have won the prize}
\label{osquebns}

There are several people who were associated with the concept of a
black hole and its development and, due to the importance of their
contributions, could have been awarded with a Nobel prize for black
holes. We now mention these people with a description of their
contributions.

Oppenheimer was born in 1904 in New York. He obtained his PhD in
Göttingen, publishing important works with Born, a German physicist
and a founder of quantum mechanics. He was professor in Berkeley and
Caltech, and from 1942 to 1945, was director of the Manhattan project
in Los Alamos that constructed the first two atomic bombs. Oppenheimer
discovered black holes together with Snyder. In a 1939 paper,
Oppenheimer and Volkoff, a PhD student in Berkeley, showed that a
neutron star, a star that had been postulated by Zwicky and 
to some extent also by
Landau, had a maximum mass of the order of the mass that Chandrasekhar
had found for white dwarfs, above which the star would collapse.
Intrigued by this
\begin{figure*}[h]
\centering
\includegraphics[scale=0.10]{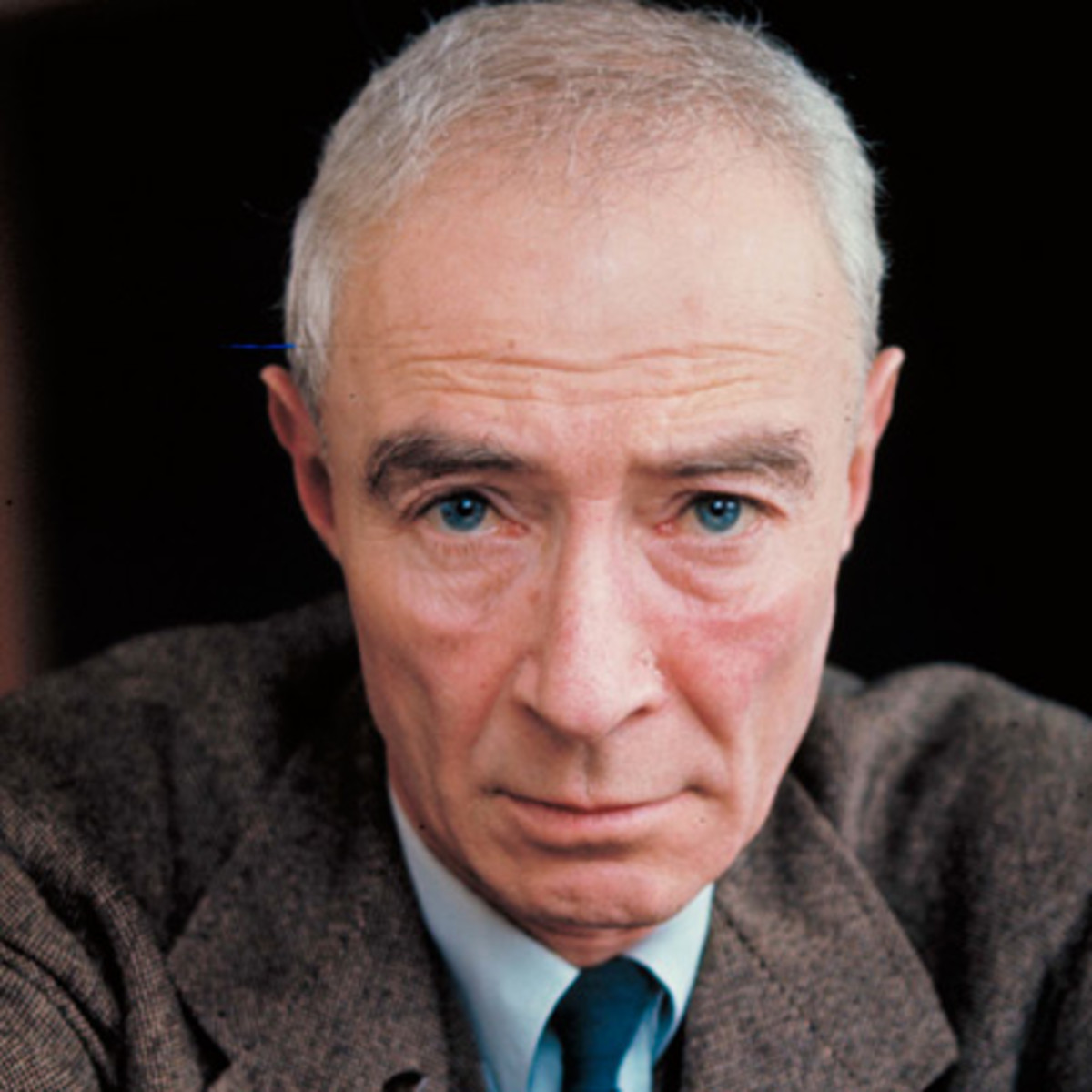}
\caption{Oppenheimer.}
\label{opp}
\end{figure*}
collapse possibility of a neutron star, Oppenheimer recruited then
another Berkeley PhD student, Snyder, with noted mathematical skills,
to solve this problem together. They formulated an idealized model of
a star collapsing under its own weight and found that the ultimate
state of such a collapse would be a black hole with an event horizon
and a singularity. In this Openheimer and Snyder 1939 paper, all the
ingredients that involve the black hole concept appear naturally,
namely, the emergence of an event horizon, beyond which the spacetime
is cut from the rest of the universe, and the description of the
distinction between two totally different times, one time marked by
the clocks of observers entering the black hole itself and running
into the singularity, another time marked by clocks of stationary
observers in the exterior that register an infinite time for the
star's surface to reach the event horizon. With his move to Los
Alamos, the political posterior incidents in the McCarthy era that
marked his life, the directorship of the Institute of Advanced Studies
in Princeton that occupied his daily routine, and the appearance of
new interests, Oppenheimer did not work in collapsed stars and black
holes ever again.  Although he knew he discovered them, he never
regarded his discovery of special worth, he seemingly considered it to
be a minor discovery. Yet as was unravelled later, the black hole is
the object par excellence of general relativity, it is pure and
complex spacetime geometry. It would have been a Nobel prize for
gravitation. Having died in 1967, he did not live to see the
extraordinary later developments from the theory and observation of
the object that he together with Snyder had created. In this initial
stage, the requirements needed for black holes and their discoverers
to be candidates to the Nobel prize in physics, let alone receive it,
were not yet satisfied.

Wheeler was born in 1911 in Florida. He obtained his PhD at John
Hopkins University
\begin{figure*}[h]
\centering
\includegraphics[scale=0.60]{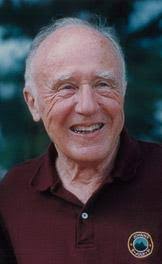}
\caption{Wheeler.}
\label{wheel}
\end{figure*}
and was professor in Princeton. He worked with Bohr in nuclear physics
and invented the S-matrix formalism to explain the quantum scattering
of particles. For that he was indicated for the Nobel prize, never
having received one. In the 1950s, he developed the physics of
collapsed stars until he was convinced that the total collapse of a
star into a black hole was inevitable, completing in a conclusive and
decisive way the Oppenheimer-Snyder collapse result. With his vision,
his PhD students created groundbreaking ideas and in certain cases
extraordinary ones, namely his student Bekenstein, born in Mexico,
answering a question of Wheeler himself correctly proposed that the
entropy of a black hole is equal to its area in appropriate units. In
1968, Wheeler decided that the name black hole was the ideal and
perfect name for completely gravitationally collapsed objects, and
later invented the expression  black holes have no hair to
synthesize that a black hole in general relativity is characterized
only by its mass, its angular momentum, and its electric charge, it
only has three hairs. Clearly someone or something with three hairs is
effectively bald. One can understand that the Academy has not given
him the Nobel prize in this area. Wheeler died in 2008, seven years
before the gravitational wave antennas confirmed beyond any doubt the
existence of black holes. In addition, Wheeler's contribution is
spread along four decades in conjunction with many students, they
themselves providing new and significant inputs. It would have been a
Nobel prize for gravitation.

Lynden-Bell was born in 1935 in Dover, England. He did his PhD with
Mestel, a British astrophysicist, in galactic dynamics, and was a
professor in Cambridge. In 1969 he made the audacious proposal,
accompanied with precise calculations to substantiate it, that all
galaxies were dead quasars, i.e., all galaxies contained a
supermassive black hole in their centers, 
although these galactic centers would have no activity for lack 
of gas and matter. Just after,
with Rees, astrophysicist from Cambridge, they showed as a corollary
of the original proposal, that our galaxy should contained a central
supermassive black hole with 4 million solar masses. Towards the
end of 1990 it was ratified that all, or almost all, galaxies have in
fact a central supermassive black hole. The spectacular image of the
black hole of the galaxy M87 done in 2018 by the EHT, acronym for
Event Horizon Telescope, is but one example.
\begin{figure*}[h]
\centering
\includegraphics[scale=0.60]{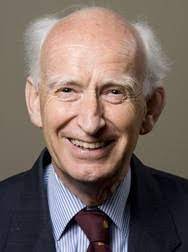}
\caption{Lynden-Bell.}
\label{dlb}
\end{figure*}
For having foreseen that all galaxies contained a central black hole,
Lynden-Bell was awarded the inaugural Kavli prize for astrophysics in
2008, together with Schmidt, an American astrophysicist of Dutch
origin from Caltech, that had identified 
optically the first quasar in 1963.
Clearly, by his daring proposal he could have been also considered as
a candidate for the Nobel prize.  The corollary of the idea thar our
galaxy contains a supermassive black hole has been proven by the
observational work of Genzel and Ghez, that won the 2020 Nobel prize.
Not having had the opportunity to celebrate this nomination, he died
in 2018, it remains for us to think that he would have been exultant
to see his ideas confirmed and cited in the announcement and in the
official ceremonies of the Academy. Lynden-Bell worked and developed
several and important topics in astrophysics and general
relativity. It would have been a Novel prize for astrophysics and
gravitation.

Hawking was born in 1942 in Oxford.  He obtained his PhD in 1966 in
cosmology and was Lucasian Professor in Cambridge, the chair that had
also been occupied by Newton and Dirac.  From 1970 onwards he started
his connection to the study of black holes to which he gave remarkable
advances. He proved rigorously within general relativity
\begin{figure*}[h]
\centering
\includegraphics[scale=0.10]{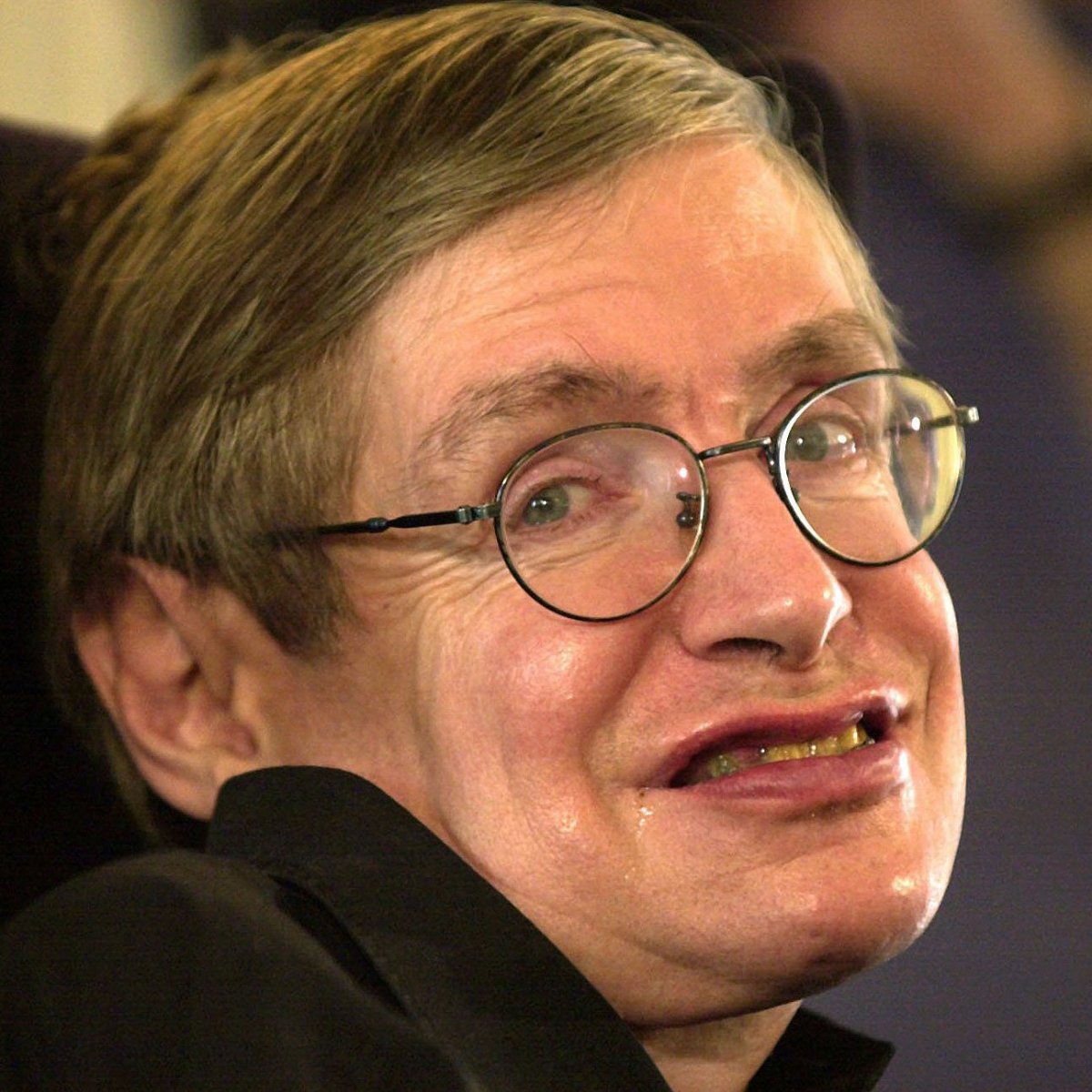}
\caption{Hawking.}
\label{hawk}
\end{figure*}
that the area of a black hole never decreases, paving the way to the
concept of black hole entropy, discovered immediately after by
Bekenstein when he was doing his PhD in Princeton and developed by
Bekenstein himself when he settled later in Israel. With Bardeen, an
American physicist, and Carter, his colleague in Cambridge, Hawking
showed that the mechanics of black holes was ruled by laws that were
identical in formal content to the laws of thermodynamics.  Combining
general relativity with quantum mechanics, two disciplines practically
immiscible, he showed in 1974 that a black hole is not black, but
irradiates at a temperature inversionally proportional to its
mass. This temperature is called Hawking temperature and its
expression involves the three fundamental constants of physics,
namely, the universal gravitational constant, the velocity of light,
and the Planck constant.  This revolutionary discovery is probably the
most important discovery in theoretical physics in the second half of
the 20th century.  With this calculation, it was definitively shown
that black holes are not only mechanical objects but also
thermodynamic objects. Now, thermodynamics is a convenient description
of nature but not a fundamental one, giving rise to the open problem
of knowing what are and where are the degrees of freedom of a black
hole that give rise to its entropy. Although these are aspects that
have been intensely studied since then and there are clues to
understanding the black hole as a quantum object, there are no
definite results. Hawking has also provided decisive contributions to
cosmology, and made efforts, through the publication of books, to
educate the general public in physics and science.  He was
a celebrity worldwide, comparable only to Einstein.  Certainly he more
than deserved the Nobel prize for all his revolutionary and
groundbreaking work in black holes. He missed receiving it by a
whisker, having died in 2018, facilitating the choice to the Academy
that can only give to three people at most. It would have been a Nobel
prize for gravitation.

\section{
From the dawn of science to the next Nobel
prizes in astrophysics and gravitation}
\label{aurora}

\subsection{From the dawn of science to black holes}
\label{aurorabns}

Anaximander, Greek philosopher that lived around 550 BC, proposed that
the ``Earth is not sustained by anything, but remains motionless due
to the fact that is equidistant to all things. Its form is that of a
drum.  We walk in one of its flat surfaces, whereas the other surface
is on the other side''.  This is considered, notably by Popper, an
Austrian science philosopher that settled in England, one of the most
portentous ideas in the entire history of human thought. To think the
Earth floating in equilibrium in the middle of space, staying still
due to equidistance, is an extraordinary abstraction, one that
anticipates Newton's idea of an immaterial force of gravitation. Of
course, the Earth is not a drum, i.e., a cylinder, the Earth is
approximately spherical, but in a cosmological context, which is the
context of Anaximader's proposal, the manuscript incorrection is
irrelevant, in fact a cylinder and a sphere are topological
equivalents in current language.

Today, 2500 years after, we have a science of astrophysics and
gravitation that explains planets, stars, galaxies, large scale
structure, the Universe itself, and black holes.  In this endeavor of
formidable discoveries all of physics is involved: general relativity
and gravitation, astrophysics, condensed matter physics, and
molecular, atomic, nuclear, and elementary particle quantum mechanics.

The appearance of the concept of a black hole as the object par
excellence of general relativity and its observational confirmation as
a naturally occurring object stands up between the great discoveries
in physics.  Decades of enthusiasm and commitment were necessary to
better understand the essence of a black hole and with that obtain a
more perfected vision of the world's structure. This discovery has,
finally and at the right time, been crowned with the 2020 Nobel prize
by the Royal Swedish Academy of Sciences in the names of Penrose,
Genzel, and Ghez.

Naturally, there are still many open problems in black hole theory and
observation.  In black hole theory, there is a fundamental problem of
understanding the interior of a black hole, the region of spacetime
that is inside the event horizon.  General relativity predicts that
inside the event horizon one finds a singularity where spacetime
itself and the laws of physics end. Related to this problem is cosmic
censorship, a hypothesis suggested by Penrose in 1969, that all
singularities are hidden inside the event horizon, i.e., there are no
naked singularities.  Still in black hole theory, there is no solution
to the major problem raised by Bekenstein, namely, since the black
hole has entropy, what are the basic constituents of that entropy, and
yet another important conundrum introduced by Hawking, namely, to know
if in the collapsing and evaporation process of a black hole, all the
information that one has in the beginning is equal to the information
one has in the end, or if not, whether the singularity has eliminated
part of that information reducing the phase space drastically.
Presumably, we will need a definitive quantum theory of gravity to
solve and explain these phenomena. From the observational side
regarding black holes, with the recent access to the gravitational
wave astronomical window joining the detection in all the
electromagnetic spectrum in all types of telescopes and detectors, one
expects that many more black holes, stellar as well as galactic,
isolated or in binary systems, distant one from the other or in
collision, will be observed. Thus, we will have a better knowledge of
the origin and evolution of such objects, of their physical
properties, and of their populational distribution; the latter could
even indicate whether black holes are important contributors to dark
matter. The observations can even test the Kerr hypothesis, that
states that astrophysical black holes are characterized by the Kerr
solution, the only features they have being their mass and angular
momentum, and also validate Hawking's black hole area theorem, which
implies that in a collision between two black holes, the area of the
final black hole is greater than the sum of the areas of the initial
black holes.  In addition, there is also the unlikely possibility of
observing very compact objects, with quasiblack 
holes or even wormholes being among those. For sure, other
problems will come along.

\subsection{The next Nobel prizes in astrophysics and gravitation}
\label{prox}

General relativity has changed our vision of the world. The fundamental
equation of the theory is
\begin{equation}
G_{ab}=\frac{8\pi G}{c^4}T_{ab}\,,
\end{equation}
where $G_{ab}$ is the Einstein tensor and is related to geometry,
$T_{ab}$ is the energy-momentum tensor and is related to matter, $G$
is the universal Newton constant of gravitation, and $c$ is the
velocity of light. In words, the Einstein equation says that geometry
and matter are coupled together. General relativity substitutes the
field and gravitational force of Newtonian gravitation for the metric
and spacetime geometry. In the wise words of Wheeler, space tells
matter how to move and matter tells spacetime how to curve.

General relativity opened major and completely new branches in
physics.  They are:
1. Gravitational lenses, foreseen by Einstein in 1912 and that
currently are an essential tool for probing the Universe;
2. Gravitational waves, worked out by Einstein in 1916 and 1918 and by
Eddington in 1922, and followed up by many;
3. Cosmology, initiated by Einstein in 1917, continued by de Sitter,
Lema\^itre, Gamow, and by many others up to the current model of
extraordinary breadth;
4. Fundamental theories, an idea formulated by Weyl in 1918 to unify
gravitation and electromagnetism and that survived until today in more
complex forms.
5. Black holes, a concept originated by Oppenheimer and Snyder in 1939
and continued by Wheeler, Penrose, Hawking, and many others.

The second, third, and fifth branches already have Nobel prizes and
certainly will have more. Gravitational waves generated in the
primordial universe, possibly in an inflationary epoch, can also be
detected in the future, which would be an admirable discovery that
would merit a Nobel prize. Cosmological discoveries, for instance the
nature of dark matter, are expected to happen sooner or later and
certainly are of Nobel prize level. The discovery of new phenomena
involving black holes, for example the collision of supermassive black
holes, will also merit the prize. The first branch, gravitational
lenses, and the fourth branch, unified theories, have not received the
Nobel prize but are destined to get it in the future. 
In fact,
gravitational lenses are a tool to probe the Universe and, 
specifically, 
to better
understand the character of dark matter. 
The importance of this lensing comes from the fact
that dark matter does
not interact through the usual electromagnetic processes with baryonic
matter of which we are made, but interacts gravitationally, so that in
astronomical concentrations,  incident light rays are bent 
to produce an effect of gravitational lens. 
A better understanding of
dark matter through gravitational lensing could be revolutionary and
thus such a discovery would also merit a Nobel prize. As theories of
unification are concerned, in their current form, their importance
lays in finding the gravitational theory that will supplant general
relativity. General relativity is a remarkable theory that applies to
scales larger than the Planck scale, where the Planck scale is defined
as the scale where quantum effects of gravitation are important,
Planck's length is $10^{-33}$cm. Theoretically there are indications
that in scales near the Planck ones, general relativity suffers
modifications, new terms related to the spacetime curvature become
important and new physical fields arise. It is thought that the big
bang, and so the Universe as a whole, emerged from those Planck
scales. This means that, since gravitation is modified at the Planck
scale, that modification must also appear at the maximum scale, the
scale of the Universe. Firm evidence that points to the correct,
new, and elusive theory of gravitation can lead to a Nobel prize.

The criteria established by the Nobel committee and by the Royal
Swedish Academy of Sciences to award a Nobel prize are very tight and
appear extremely correct. This is evident from the incalculable
prestige of the Nobel prize in physics. Let us examine two
criteria. The criterion that the awarded work has to have experimental
confirmation, is an appropriate criterion given that physics is a
science of nature.  Theoretical speculations are interesting and
important, but they only become physics in a proper sense when they
have confirmation. The other criterion, that the prize is given to at
most three people, is very important. In this way it is guaranteed
that the selection process is thorough, accurate, and robust, avoiding
a devaluation of the prize because of dispersion. There are theses
that the prize could also be given to big collaborations, but that
would be inadequate. Imagine giving the prize to a scientific
collaboration of 500 people. This means that each person of the
collaboration has 1/500 of the prize, and clearly could not be
considered effectively holder of a Nobel prize. One cannot mix an
individual prize with a prize for an institution as it is the case of
a collaboration. It would be propitious that some organization could
create a special prize to big collaborations in science, encompassing
physics, chemistry, biology, physiology, and possible other scientific
disciplines, but the Nobel prize in physics as we know it is
individual, or better, to three people at maximum, a fact that
contributes to the great prestige that the prize has. Finally, when
one awards a prize, in particular a Nobel prize, one has to make
choices, and the choices always have a political side,
science being no exception. 
For example, one needs to choose to which area of physics the
prize will go in that year, or to ascertain if there is an outstanding
discovery that supersedes the choice of the area. The detection of
gravitational waves enters clearly in the latter case, they were
detected in 2015 and the Nobel prize was awarded one year and a half
later, in 2017. When one has to select three names out of several,
there are those that stay outside.  Nevertheless, the Nobel committee
has been nearly flawless in their decisions. One can always argue that
this or that name should have received the prize, in this context the
name of Sommerfeld, a great
German physicist that never received the prize,
invariably comes up, but as a whole, after more than one century of
Nobel prizes in physics, the choices have been consistently
correct. We can thus easily understand why the Nobel prize in physics
has maximum prestige. Those who receive it have to be outstanding,
clearly. But, in the midst of those outstanding afortunates, one has
to distinguish those that yield prestige to the prize and those that
receive prestige from the prize.  Now, practically all the great names
in physics have received the Nobel prize, and being they the great
names of physics, confer in turn to the prize the enormous prestige
that it has. The Nobel prize in physics is, by a great margin, the
most important prize in physics, one can even say that it is the only
prize of interest at a planetary scale. As it should have been made
clear, the physics Nobel prize, beyond a prize for physicist, is a
prize for physics as a science and a human endeavor.

We count with truly extraordinary new discoveries in astrophysics and
gravitation, and in particular in gravitational waves and black holes,
in a way as to merit, amidst the rigorous evaluations of the committee
and the Academy, many more Nobel prizes. This would be motif of pride
and glory, not properly by the prizes in themselves, but by what they
represent and mean, notably, that our physical understanding of the
Universe enlarges ever more.

\section*{Acknowledgements and notes}
\label{agrad}

Acknowledgements:
I thank Carlos Herdeiro for the collaboration as a restless editor in
the edition of the special number of Gazeta de Física on black holes
with several authors commemorating the 2020 Nobel prize in physics.  
I
thank Bernardo Almeida, director of Gazeta de Física, for liking the
idea of having this special edition. 
I thank Fundação para a Ciência e
Tecnologia - FCT, Portugal, for financial support through project
UIDB/00099/2020.

Notes: Text in inverse commas means the transcription of the Nobel
committee statement.  This article is a concomitant translation to
English of the article written in Portuguese and published in Gazeta
de F\'isica 44(2/3), 58 (2021). I very much 
thank Justin Feng for the great help in 
putting a preliminary rough translation of the article 
into a fluent Saxonic version.

\section*{References}\label{bib}

\noindent
[1] C. A. R. Herdeiro and J. P. S. Lemos, ``The black hole fifty years
after: Genesis of the name'', arXiv:1811.06587 (2018), translation of
the original in Portuguese, C. A. R. Herdeiro, J. P. S. Lemos, ``O
buraco negro cinquenta anos depois: A génese de um nome'', Gazeta de
Física 41(2), 2 (2018).

\noindent
[2] The Nobel Committee for Physics, ``Scientific Background on the
Nobel Prize in Physics 2020: Theoretical foundations for black holes
and the supermassive compact object at the galactic centre'', Advanced
Information Kungliga Vetenskapsakademien (2020),
https://www.nobelprize.org/prizes/physics/2020/advanced-information/

\noindent
[3] The Nobel Prize, ``All Nobel Prizes in Physics'' (2021),
https://www.nobelprize.org/
prizes/lists/all-nobel-prizes-in-physics/.

\end{document}